\documentclass[12pt]{article}
\usepackage{array}
\usepackage{tabularx}
\usepackage{amsmath,amssymb,amsfonts,mathrsfs,bbm,epsfig,cancel,cite,setspace,bigstrut,framed,eufrak,tensor,mathtools}
\usepackage[all]{xy}
\usepackage{color}
\usepackage{pifont}
\usepackage{xcolor}
\usepackage{tikz}
\usepackage{mathrsfs}
\usetikzlibrary{positioning}

\usepackage{mathtools}

\newcommand{\dsl}{\pa \kern-0.5em /}

\newcommand{\pa}{\partial}

\newcommand{\bit}{\begin{itemize}}
\newcommand{\eit}{\end{itemize}}

\def\bZ{\mathbb Z}

\newcommand{\bea}{\begin{eqnarray}}
\newcommand{\eea}{\end{eqnarray}}
\newcommand{\be}{\begin{equation}}
\newcommand{\ee}{\end{equation}}

\newcommand{\ba}{\begin{array}}
\newcommand{\ea}{\end{array}}
\def\bZ{\mathbb Z}

\makeatletter \@addtoreset{equation}{section} \makeatother

\author{JQ}

\newcommand{\comment}[1]{}

\newcommand{\bR}{{\mathbf R}}

\newcommand{\ks}{{\mathfrak s}}
\newcommand{\ko}{{\mathfrak o}}
\newcommand{\ku}{{\mathfrak u}}

\newcommand{\kg}{{\mathfrak g}}
\newcommand{\kp}{{\mathfrak p}}

\newcommand{\kso}{{\ks\ko}}
\newcommand{\ksp}{{\ks\kp}}
\newcommand{\ksu}{{\ks\ku}}

\newcommand{\cI}{{\cal I}}

\newcommand{\cT}{{\cal T}}
\newcommand{\cZ}{{\cal Z}}

\newcommand{\cD}{{\cal D}}

\newcommand{\cF}{{\cal F}}
\newcommand{\cA}{{\cal A}}

\newcommand{\iZ}{\mathbbm{Z}}

\newcommand{\ii}{\mathbbm{i}}

\newcommand{\sD}{\mathscr{D}}

\newcommand{\bi}{\begin{itemize}}
\newcommand{\ei}{\end{itemize}}
\newcommand{\beq}{\begin{equation}}
\newcommand{\eeq}{\end{equation}}

\newtheorem{theorem}{\bf THEOREM}

\begin{document}


\begin{titlepage}
\vfill
\begin{flushright}
{\tt\normalsize KIAS-P24022}\\

\end{flushright}
\vfill
\begin{center}
{\Large\bf Discrete Gauge Anomalies and Instantons}

\vskip 1.5cm

Qiang Jia and Piljin Yi
\vskip 5mm

{\it School of Physics,
Korea Institute for Advanced Study, Seoul 02455, Korea}

\end{center}
\vfill

\begin{abstract}
We revisit anomalous phases related to large gauge transformations, such as
the Witten anomaly. The latter, known to plague $d=4$ $Sp(k)$ theories, is
well-understood in terms of $\pi_4(Sp(k))=\iZ_2$, but it also has an oblique
relation to the instantons, labeled by $\pi_3(G)=\iZ$, via the fermion
zero mode counting. We revisit this relation and point out
how $SU(N)$ theories escape an anomalous sign of the latter type,
only thanks to the perturbative anomaly cancelation condition that restricts the
chiral fermion spectrum. This leads to the question of what happens if
the latter, more mundane anomaly is canceled by an inflow instead. After
raising an open question about fractional D3 probe theories, 
we explore the simplest bottom-up model of such a kind, due to Witten and Yonekura,
from which we find the relevant chiral theories to be free of such a disease despite the
unrestricted chiral spectra. We close with a simple but often-overlooked
observation about how fermionic zero modes enter physics differently between
Euclidean and Lorentzian descriptions and point out a related issue in $d=3$.

\end{abstract}

\vfill
\end{titlepage}

\tableofcontents

\section{$Sp(k)$ vs. $SU(N)$ and Dynkin Indices}

Much of this paper is about the differences and similarities between
$Sp$ and $SU$ gauge theories. The two classical Lie groups are nominally
very different from each other. For instance, most irreducible
representations of $SU(N)$ theories are complex while their counterpart
for $Sp(k)$ are real or pseudo-real, so one can hardly say that the two are
similar. In $d=4$ another well-known difference is how the $Sp(k)$ theories
are automatically anomaly-free under small gauge transformations while
$SU(N)$ theories are free of the Witten anomaly associated with $\pi_4(G)$
of the gauge group\cite{Witten:1982fp}.

On the other hand, some common properties set them
apart from other Lie Groups in the nonperturbative gauge dynamics. Recall
that  $\pi_3(G)=\iZ$ for all $G$.
The winding number jump of the latter between the past infinity and the future
infinity famously involves the Yang-Mills instanton\cite{Belavin:1975fg, Callan:1976je, Jackiw:1976pf}. The unit instanton
of $SU(2)=Sp(1)$ can be embedded into $SU(N)$ and $Sp(k)$ naturally, hinting
at a common thread between the two. The similarity between the two in this
context becomes quite stark when we compare this to how Yang-Mills instanton
behaves when embedded into $SO(N\ge 7)$ theories.

For the latter, the smallest possible such embedding is done via
\bea
\ksu(2)_\pm \subset \kso(4)\subset \kso(N\ge 7)
\eea
where $\ksu(2)_\pm$ is the (anti-)self-dual subalgebra  such that
$\kso(4)=\ksu(2)_+\oplus \ksu(2)_-$. Under this, the smallest number
of zero modes around this minimal instanton is 2 for the defining
representation of $\kso(N\ge 7)$.
This should be contrasted against how its counterpart for the same
minimal instanton in $\ksu(N\ge 2)$ and $\ksp(k\ge 1)$ theories is 1
again for the defining representations.

A general statement can be made with
the Dynkin index  $D_2^\bR(\kg)$ of the representation  $\bR$ and the
gauge algebra $\kg$, defined via
\bea\label{DynkinIndex}
{\rm tr}_\bR^\kg(\cT^A\cT^B) \;\;=\;\; D_2^\bR(\kg)\times {\rm Tr}_\kg(\cT^A\cT^B)
\eea
This is still ambiguous since the representation-independent trace
$ {\rm Tr}_\kg$ can be defined in various ways for each Lie algebra;
one standard choice is to let
\bea
D_2^{\rm adj}(\kg)=2h^\vee_\kg
\eea
for the adjoint representation, with the dual Coxeter number $h^\vee_\kg$.\footnote{
The Dynkin  index can be also defined as
\bea
 D_2^{\mathbf R}(\kg) = \frac{{\rm dim}\, \mathbf R}{{\rm dim} \, \kg}\times \mu_{\mathbf R}\cdot (\mu_{\mathbf R}+2\rho_\kg)
\eea
where $\mu_{\mathbf R}$ is the highest weight of $\bR$ and $\rho_\kg$ is the Weyl vector,
which is half the sum over positive roots. The normalization ambiguity above
manifests in the normalization choice of the root system. This expression
is known to give $2h^\vee$ for the adjoint representations if the
normalization of roots is chosen such that the long roots have the
length squared equal to 2.}

With this
\bea
D_2^\bR(\kg)
\eea
is always an integer and counts the index of a fermion in the representation
$\bR$, modulo a sign, in the background of the minimal $\kg$-instanton.
Furthermore, this number is always even unless $\kg=\ksu,\ksp$. As already noted,
we find
\bea
D_2^{\rm defining}(\ksu)=    1=D_2^{\rm defining}( \ksp)\ , \qquad D_2^{\rm defining}(\kso)=   2
\eea
It should be by now clear why we confined
our comparison to $\kso(N\ge7)$; with $N\ge 6$, for which
alternate $\ksu/\ksp$ descriptions are possible, the smallest
$D_2^\bR(\kso)$ arises from (chiral) spinor representations,
which are the defining representations in the $\ksu/\ksp$ sense.

The above ${\rm Tr}_\kg$ is such that
the minimal instanton has the same instanton number 1, and thus differs
from ${\rm Tr}^\kg_{\rm defining}$ when $\kg=\kso$. The $\kso$ Yang-Mills 
instantons would be quantized in unit of 2 in the latter version of the trace.
This factor $2$ difference between $\ksu/\ksp$ and $\kso$ manifests in
many places in gauge theories where typically we write the Yang-Mills
kinetic term in the defining representation. 
One should not be fooled into thinking that this  can be
accounted for by the normalization choice for gauge generators that
we sometimes encounter early when introducing gauge theory action.
Note how the above distinction comes from the Atiyah-Singer
index counting which has nothing to  do with normalization of anything.

One of the more widely known consequences of this factor $2$
affects how the Chern-Simons coefficients are quantized, and
also in a factor-two longer periodicity of $\theta$-angle in $d=4$ $\ksu/\ksp$
theories, relative to $\kso$ theories\cite{Aharony:2013hda}.
The same $\theta$-periodicity doubling for $\ksp$ theories relative to $\kso$
theories has been noted by E. Witten decades ago in the context of the type IIB
Orientifold constructions\cite{Witten:1998xy}.\footnote{Please be
aware that in recent literature, one finds yet another type of
$\theta$-periodicity statements, which seemingly arise from the Lie
group choices rather than the Lie algebra choices. The latter has
nothing to do with our observation here; we are working in the
context of the vanilla field theory with neither external and extended
defect nor the possibility of gauging (part of) the center of $\kg$.}

More generally, $D_2^{\bR}(\kg)$ is always even unless $\kg=\ksu, \ksp$.
In this sense, there is a close affinity between $\ku$ and $\ksp$
when it comes to the instanton physics and other related topological sectors.
As mentioned in the beginning, however, the two are also very distinct when it comes to $\pi_4(Sp(k))=\iZ_2$ and $\pi_4(SU(N>2))=\emptyset$
so that only in the former  discrete gauge anomalies of the Witten type
become possible, as we review next. This brings about a conceptual
difficulty on the $\ksu$ side when we remind ourselves of a close connection between
the Witten anomaly and the instanton physics, observed already in the original
paper and used as the practical computational tool thereafter.

Although there is no real issue in the end once we impose the one-loop
perturbative anomaly cancelation at purely field theory level,
things become a little
more confusing when we begin to embed $d=4$ theories to a higher dimensional set-up
such as via geometrical engineering. Our primary aim in this note is to
point out the subtleties involved and clear up the issues as much as we can.

\vskip 5mm

N.B. The questions being raised here were formulated and developed 
as part of a graduate text on advanced quantum field theories \cite{Vol2} 
by the senior author, from which we borrowed the bulk of the review material.

\section{An Overview of the Witten Anomaly}

Let us make a brief overview of Witten's discrete anomaly for $d=4$
$Sp(k)$ gauge theories\cite{Witten:1982fp}. For this, we start with a chiral fermion in
$d=2n$  with the kinetic term written in the chiral basis
\bea
\left(\begin{array}{cc} 0 & \chi^\dagger \end{array}\right)
 \ii \gamma^a \sD_a
\left(\begin{array}{c} \psi \\ 0\end{array}\right) =
\left(\begin{array}{cc} 0 & \chi^\dagger \end{array}\right)
\left(\begin{array}{cc} 0 & \ii \bar\sigma^a\bar \cD_a  \\
\ii \sigma^a \cD_a & 0\end{array}\right)
\left(\begin{array}{c} \psi \\ 0\end{array}\right)
\eea
with $\bar\sigma^a=(\sigma^a)^\dagger$ and $\bar\sigma^a\sigma^b +
\bar\sigma^a \sigma^b=2\delta^{ab}$ and vice versa, in the Euclidean
signature. We will denote the spacetime manifold $Y_{2n}$.

Although there is no eigenvalue problem here, $ (\ii \gamma^a \sD_a)^2$ on $\psi$
does admit eigenvalues which we call $\lambda^2$. One way to deal with the
above kinetic action is to extend the spinor artificially as
\bea
 \ii \gamma^a \sD_a
\left(\begin{array}{c}  \psi_\lambda  \\ \pm \ii \sigma^a\cD_a \psi_\lambda /\sqrt{\lambda^2}  \end{array}\right)
= \pm \sqrt{\lambda^2}
\left(\begin{array}{c} \psi_\lambda  \\ \pm \ii\sigma^a \cD_a  \psi_\lambda /\sqrt{\lambda^2}  \end{array}\right)
\eea
which is consistent with the fact that in even dimensions the eigenvalues $\lambda$
of the Dirac operator come in pairs with mutually opposite signs. Since
the lower half of these extended spinors is not physical,  we  pick one of
the two, either $\sqrt{\lambda^2}$ or $-\sqrt{\lambda^2}$ leading to  the Pfaffian,
\bea
\cZ_Y=
{\rm Pfaff}( \ii \gamma^a \sD_a) = \prod_{\lambda>0}\lambda
\eea
where we assumed that no zero mode exists on $Y$.

The Witten anomaly arises because the above restriction to one
sign of the pairwise eigenvalues of $ \ii \gamma^a \sD_a$ on Dirac spinor
can be ambiguous. For this, imagine a fictitious $d=2n+1$
dimensional manifold $\mathbbm{Y}_{2n+1}$ with the topology,
$[0,1]\times Y_{2n}$. We will let $Y_{2n}$ to change continuously with $x\in[0,1]$,
and keep track of how eigenvalues $\lambda_{(x)}$ evolves between
$Y_{2n}^{(x=0)}$ and $Y_{2n}^{(x=1)}$.
Here we have two choices for defining the Pfaffian. One is to maintain
the above definition of the Pfaffian at each and every $x$,
\bea
{\rm Pfaff}( \ii \gamma^a \sD_a)\biggr\vert_x = \prod_{\lambda_{(x)} >0}\lambda_{(x)}
\eea
or, instead, define
\bea
\widehat {\rm Pfaff}( \ii \gamma^a \sD_a)\biggr\vert_{x} = \prod_{\lambda_{(0)} >0}\lambda_{(x)}
\eea
which continuously follow the individual eigenvalues that are positive
at $x=0$. Both definitions look perfectly sensible, and in fact, in most
situations, the two would agree with each other.

Suppose that, for some $\mathbbm{Y}$, $N$-many positive (negative) eigenvalue
$\lambda_{(0)}>0$ crosses zero over to the negative (positive) side,  as we
follow $x\in[0,1]$. We then find
\bea
\widehat{\rm Pfaff}( \ii \gamma^a \sD_a)\biggr\vert_{x=1} =(-1)^N\,{\rm Pfaff}( \ii \gamma^a \sD_a)\biggr\vert_{x=1}
\eea
The two equally sensible definitions collide badly and even
$\widehat{\rm Pfaff}$ can collide with itself since the interpolation
$\mathbbm{Y}_{2n+1}$ need not be unique.

\begin{figure}[!h]
    \centering
    \includegraphics[scale=0.6]{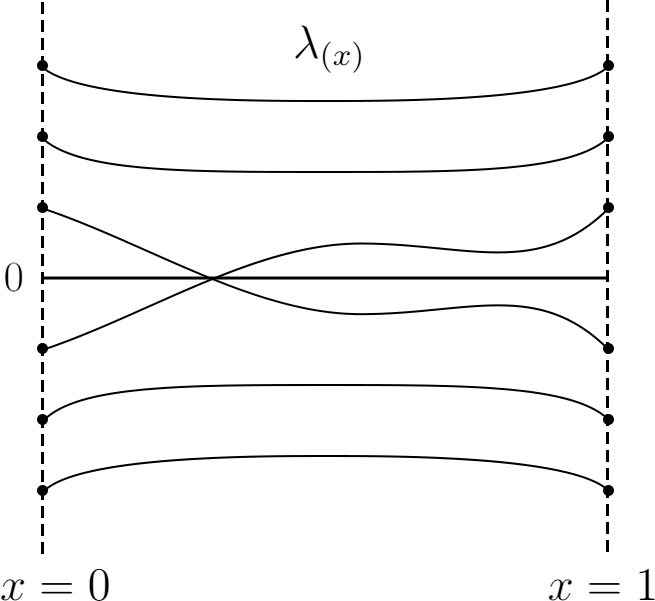}
    \caption{Eigenvalue crossing}
    \label{fig:eigenvalue-crossing}
\end{figure}

This becomes a matter of
internal consistency if $Y_{2n}^{(x=0)}$ and $Y_{2n}^{(x=1)}$ are
related by a gauge transformation.
If the gauge transformation is ``small," $N=0$ since eigenvalue does
not change under any gauge transformations. If the gauge transformation
between  $Y^{(1)}_{2n}$ and $Y^{(0)}_{2n}$ is ``large,"
the intermediate $Y^{(x)}_{2n}$ at generic $x$ cannot be gauge equivalent to $Y^{(1)}_{2n}$ or
to $Y^{(0)}_{2n}$ so that $N\neq 0$ is possible. In the latter situation,
a self-consistency problem arises if $N$ of any such $\mathbbm{Y}$ is odd, since
\bea
\widehat{\rm Pfaff}( \ii \gamma^a \sD_a)\biggr\vert_{x=1} =-\,{\rm Pfaff}( \ii \gamma^a \sD_a)\biggr\vert_{x=1}
\eea

For this kind of inconsistency to occur, a necessary condition is the existence
of a large gauge transformation, which is classified by the homotopy group\cite{hu1959homotopy,toda1963composition},
\bea
\pi_d(G)
\eea
Witten's original observation comes with
\bea
\pi_4(G) = 0
\eea
for most simple Lie Groups, except
\bea
\pi_4(Sp(k)) = \iZ_2\ .
\eea
So this type of anomaly is possible in $d=4$ gauge theories only for $Sp(k)$.
The question comes down to the following: for a $d=4$
$Sp(k)$ gauge theory with a single chiral fermion $\psi_{\mathbf R}$
in some irreducible representation $\mathbf R$ of $Sp(k)$, when do
we have $(-1)^N=-1$ for $\mathbbm{Y}_5$ interpolating between a pair
of $Y_4$'s mutually related by a large gauge transformation. Since
$\pi_4(Sp(k)) = \iZ_2$ with its single non-trivial element, it is
a matter of asking for what $\mathbf R$, we have $(-1)^N =-1$ under
the non-trivial large gauge transformation in $\pi_4(Sp(k))$.

\subsubsection*{Mapping Torus and Mod-2 Index}

We can solidify this a little more by gluing the two boundaries
of $\mathbbm{Y}_{d+1}$, namely $Y_d^{(0)}$ and $Y_d^{(1)}$. These two are related
by a gauge transformation, even though the latter is not continuously connected
to the identity, so should be considered the same.
The resulting compact manifold, $\hat{\mathbbm{Y}}_{d+1}$, is called the mapping
torus. The question of $(-1)^N=\pm 1$ translates to whether the mapping torus
accepts an even or odd number of the zero modes and whether this mod 2 counting
of zero modes is topologically protected\cite{Atiyah:1976qjr}.
For the immediate problem of $d=4$ $Sp(k)$ Witten anomaly, this allows
the mod-2 index in $d+1=5$ dimensions.

The Dirac equation on the bulk $\hat{\mathbbm{Y}}_{d+1}$, with $x$-dependence
considered to be slow, would be
\bea
\ii \boldsymbol\gamma^x \left(\partial_x+\mathbb D_{(x)}\right)\Psi=0\ ,
\qquad \mathbb D_{(x)}\equiv\boldsymbol\gamma^x \gamma^a{\sD}_a\biggr\vert_x
\eea
which is similar to the zero mode equation we wrote for the APS index,
except that the odd and the even dimensions flipped the roles. Here,
$\boldsymbol\gamma^x$ is the last $\gamma^{2n+1}$ on $\hat{\mathbbm{Y}}_{2n+1}$,
which  in turn may be used as a chirality operator $\Gamma$ on $Y_{d=2n}$. $\mathbb D_{(x)}$ behaves much
like the Dirac operator $Y_d^{(x)}$, in that it is hermitian and also anticommutes
with $\boldsymbol\gamma^x$.

Although $\mathbb D_{(x)}$ is not quite the same as the usual $d$-dimensional Dirac operator,
$\ii\gamma^a\sD_a$ at $x$, the two share the same set of eigenvalues and the
same chiral building blocks for the eigenfunctions.
 These can be seen most clearly in the chiral basis we have adopted on and off.
In the end, this means that we again have a pair of eigenmodes,
\bea
\mathbb D_{(x)}\Psi_{(x)}^{(\lambda)}  = \lambda\Psi_{(x)}^{(\lambda)}\ ,
\qquad \boldsymbol\gamma^x\Psi_{(x)}^{(\lambda)} =\Psi_{(x)}^{(- \lambda)}
\eea
One should note that $\Psi_{(x)}$ came from a Dirac spinor $\Psi$ on $\mathbbm{Y}_{d+1}$,
so is a Dirac spinor on $Y_d$, while
the physical spinor $\psi$ was a Weyl spinor on $Y_d$.

A single ``eigenvalue" crossing of $\psi$ really corresponds to an eigenvalue crossing
for a pair $(\lambda,-\lambda)$ of $\Psi_{(x)}$. Do we still have the statement that each eigenvalue
crossing generates one zero mode on the mapping torus? The answer is yes.
Given such a crossing of a  $\lambda_{(x)}$ eigenmode, The $x$-dependence of $\Psi$
would be, approximately,
\bea
\Psi\; \simeq \; e^{-\int_{x_0}^x \lambda_{(x)}dx} \times \Psi_{(x_0)}^{(\lambda_{(x_0)})}
\eea
where $x_0$ denotes the location of the eigenvalue crossing,
and normalizable mode would emerge only for the eigenvalue that crosses
from negative to positive as we move along the increasing direction of $x$.
Therefore, each eigenvalue crossing of the pair $(\lambda,-\lambda)$ on $Y_d$
indeed implies a single zero mode on $\mathbbm{Y}_{d+1}$.

On the other hand, for $d=8k+2, 8k+4$, depending on whether the gauge representation
is real or not, we have the following pattern of repetitions of eigenvalues:
\begin{itemize}
\item If real, $(\lambda,\lambda,-\lambda,-\lambda)$.
\item If  pseudo-real or complex,  $(\lambda,-\lambda)$.
\end{itemize}
Although the relevant eigenvalue crossing is that of $\mathbb D$ rather
than $\ii \gamma^a\sD_a$, the repetition pattern remains the same in
all even dimensions, which can be traced to the fact that the eigenstates
of two operators came from the same chiral building blocks. For $d=8k, 8k+6$,
real and pseudo-real are swapped.

Coming back to the $d=4$ problem at hand, we see that $d=4$ eigenvalue crossings
are automatically doubled for real gauge representation. Such an event cannot
flip the sign of the Pfaffian of the $d$-dimensional Dirac operator, so no
anomalous sign problem can arise. This applies to any tensor of $SO$ vector
representations, including the adjoint. If pseudo-real,
such as for the defining representation of $Sp(k)$, or complex as in the defining
representation of $SU(N)$, there is a logical possibility of an odd number
of eigenvalue-crossing, and thus of the discrete gauge anomaly.

This gives us yet another reason to pay attention to $Sp$ gauge groups,
but the same also tells us to be wary of $SU(N)$ theories. The eigenvalue
crossing alone seems to suggest that $SU$ theories are similar to $Sp$ theories,
rather than to $SO$ theories. The bulk of this paper is to understand how $SU(N)$
theories evade the Witten anomaly from various viewpoints.
On face value, one might be content with $\pi_4(SU(N))=\emptyset$
and be done with it, but things become a little more subtle with
$SU(N)$ theories because of a purported connection between the Witten
anomaly and the instanton zero mode counting that we review next.

\section{Witten Anomaly vs. Instanton Zero Modes}

Although this mod 2 index on $\hat{\mathbbm{Y}}_5$, or equivalently
the mod-2 counting of the eigenvalue-crossing, is a well-defined topological
quantity, the actual computation is at best cumbersome. Fortunately,
there is a different litmus test for the Witten anomaly, curiously
based on the instantons, classified by $\pi_3(G)$ rather than by $\pi_4(G)$.
The claim is that the discrete $\pi_4(G)$ gauge anomaly
occurs if and only if a unit instanton admits an odd number of
fermionic zero modes. Let us first see exactly how this
relation comes about.

Given the Atiyah-Singer index theorem, the latter
fermion zero modes would be counted by the sum
\bea
\sum_{\mathbf R} \pm \frac{1}{8\pi^2}\int_{Y_4} {\rm tr}_{\mathbf R} \cF\wedge\cF
\eea
over the representation $\mathbf R$ of the chiral fermions, with
$\pm$ referring to the chirality. As we have noted in the introduction,
computation of this number for a minimal instanton gives
\bea
\sum_{\mathbf R}\pm D_2^\bR(\kg)
\eea
which is a sum of Dynkin indices. The formula counts indices
rather than net zero modes, but since we are counting the net number mod 2
and since individual indices are integral, neither of these matters.

When the net fermion zero mode count is odd, the theory is clearly
inconsistent at the quantum level, since the instantons would generate
a non-perturbative saddle contribution to the renormalized action,
of a Grassmann-odd type.
One can also show how this leads to a failure of the gauge invariance
under the large gauge transformation in spatial slices, classified
by $\pi_3(G)=\iZ$, as delineated by Witten in one of his earlier papers
on the subject \cite{Witten:1985xe} and attributed to Jeffrey Goldstone.

 The line of thought goes as follows. For simplicity, take
 the 4d spacetime to be $Y = \mathbbm{S}^1_\tau\times\mathbbm{S}^3_{z}$ with $\tau
 $ the time direction. Denote $\pi$ as the operator that performs the large
 gauge transformation on the spatial slice $\mathbbm{S}^3_{z}$, a generator of
 $\pi_3(G)$. Namely, if $|\mathcal{A}(z),\Psi(z),\cdots\rangle$ is a state in Hilbert space which is an eigenstate of the space components of the fields variable $\mathcal{A},\Psi,\cdots$. Then $\pi |\mathcal{A}(z),\Psi(z),\cdots\rangle = |\mathcal{A}^g(z),\Psi^g(z),\cdots\rangle$ where $g$ stands for the generator of large gauge transformation in $\pi_3(G)$.

 Let $J$ be the generator of a rotation about an arbitrary axis in $\mathbbm{S}^3_{z}$ and consider the operator
	\begin{equation}
	G_{\varphi}^{t';t} = \pi^{-1}_{t'} \exp(-\ii \varphi J) \pi_t \exp(\ii\varphi J)
	\end{equation}
for $\varphi \in [0,2\pi]$. The operator has the
effect of $\pi$ followed by its inverse, except that we rotate the action of the first
$\pi$ rotated by an angle $\varphi$. Obviously, $G_0^{t;t}=1$ is the identity operator.
With $t'=t$ but with nontrivial $\varphi$, the discrete $\pi_3$ large gauge transformation 
cancels out, leaving behind small gauge transformations parameterized by $\varphi$.
Note how physical states obey Gauss's law and should be invariant under any small gauge
transformation, modulo the perturbative anomaly. Since $Sp(k)$  gauge theories admit no
no perturbative anomaly in $d=4$, on the other hand, this translates to $G_\varphi^{t;t}=1$ for any $\varphi$.

On the other hand, the action $\pi$ creates an instanton since the twisting
means a winding number jump between before and after. If the instanton happens to be
equipped with  $N$ fermion zero modes, the path integral is non-zero only if we
insert the $N$ fermion fields to soak the zero modes. From a Hamiltonian point of view,
it means $\pi$ creates $N$ fermions, so that $ (-1)^F\pi = (-1)^N \pi(-1)^F $.
With $N$ odd, we, therefore, end up with $G_{2\pi}^{t'>t}=-1$, which conflicts
against the  limit $G_{2\pi}^{t;t} = \pi^{-1}_t (-1)^F \pi_t (-1)^F = 1$.

An odd number of fermionic zero-modes for an instanton also 
implies that the effective vertex of the instanton process has to be equipped with 
an odd number of fermion fields. The vertex would be Grassmanian and also there appears 
to be no way to make a Lorentz-invariant combination in dimensions higher than one.
We will revisit this curiosity in the last section of this note. 

\subsection{Pontryagin-Thom and Instanton Zero Modes}

However, this does not quite tell us why this inconsistency is connected to
the Witten anomaly that we discussed above. The connection requires
a more detailed study of the non-trivial map in $\pi_4(Sp(k))=\iZ_2$,
and how one may construct such a winding configuration starting from
an instanton.

For an illustration, it is simplest to take
$Y=\mathbbm{S}^1_\tau\times\mathbbm{S}^3_{z}$ and put a unit winding map
of $\pi_3(S^3)$,
\bea
U(z):\quad \mathbbm{S}^3_{z}\quad \rightarrow\quad S^3=Sp(1)
\eea
as $2\times 2$ unitary matrix, and then twist it along
$\mathbbm{S}^1_\tau$ as
\bea \label{homotopy-map}
U(z,\tau)\equiv e^{\ii \pi \tau \sigma_3}U(z)e^{-\ii \pi \tau \sigma_3}\ , \qquad \tau\;\in\;[\,0,1)=\mathbbm{S}^1_\tau
\eea
This is essentially a Pontryagin-Thom construction,
which says that this $U(z,\tau)$ constitutes the
nontrivial element in $\pi_4(Sp(1))$.

Let's illustrate the Pontryagin-Thom construction for $\pi_4(Sp(1))$ where the group manifold of $Sp(1)$ is $S^3$. For a more detailed
description, we refer the readers to Appendix A. By Pontryagin-Thom theorem, the homotopy group $\pi_4 (S^3)$ is equivalent to the set of 1-dimensional circles $S^1$ in $S^4$ up to framed cobordism. Consider a circle $\gamma$ parametrized by $\tau\in [0,1)$. If we ignore the framing, then any curve $\gamma$ in $S^4$ is shrinkable and trivial since $H_1(S^4)=0$. The normal bundle is $\mathbb{R}^3$ and we can choose a section $\mathfrak{o}_{\gamma,\tau=0}$ of the normal bundle at $\tau=0$ and extend it over the curve $\gamma$ to construct a framing. Then the section $\mathfrak{o}_{\gamma,\tau}$ determines a map from the circle $\gamma$ to $SO(3)$ group which is classified by $\pi_1(SO(3)) = \mathbb{Z}_2$.

Therefore there are two ways to construct the framing: if the section $\mathfrak{o}_{\gamma,\tau}$ defines a trivial map in $\pi_1(SO(3)) = \mathbb{Z}_2$ then the curve $\gamma$ is shrinkable and corresponds to a trivial element in the framed cobordism group; on the other hand, if the section $\mathfrak{o}_{\gamma,\tau}$ gives the non-trivial map in $\pi_1(SO(3)) = \mathbb{Z}_2$, then one cannot shrink the circle $\gamma$ while preserving the framing. However, if we have another circle $\gamma'$ parametrized by $\tau \in [1,2]$ which also has a non-trivial framing $\mathfrak{o}_{\gamma',\tau}$, we can join $\gamma$ with $\gamma'$ to a new circle $\gamma''$ with $\tau \in [0,2]$ and the framing $\mathfrak{o}_{\gamma'',\tau}$ will be trivial. Therefore, the framed cobordism group for a 1-dimensional curve in $S^4$ is $\mathbb{Z}_2$ and by Pontryagin-Thom the homotopy group $\pi_4(S^3)$ is also $\mathbb{Z}_2$.

Then we can recover the homotopy map $U(x): S^4 \rightarrow S^3$ as follows. Pick a tubular region $\gamma \times \mathbb{R}^3_d \in S^4$ where $\mathbb{R}^3_d$ is a 3-dimensional ball with a small radius $d$ and the standard frame of $\mathbb{R}^3_d$ induces the section $\mathfrak{o}_{\gamma,\tau}$ which varies along the curve $\gamma$. The homotopy map $U(x)$ is constructed as
\begin{equation}
   U(x) = \left\{ \begin{array}{ll}
         \frac{\pi(x)}{\beta(|x|)}\quad \textrm{if}\ x \in \gamma \times \mathbb{R}_d^3 \\
          \infty\quad \textrm{if}\ x \notin \gamma \times \mathbb{R}_d^3
    \end{array}\right.
\end{equation}
where the projection map $\pi: \gamma \times \mathbb{R}^3\rightarrow \mathbb{R}^3$ and the cut-off functions $\beta(x)$ are defined in the appendix. Here the target space $S^3$ is identified with $\mathbb{R}^3 \cup \{\infty\}$ by stereographic projection. It is easy to see the map $U(z,\tau)$ defined in \eqref{homotopy-map} is a representative of the non-trivial homotopy map if we identify $\gamma \equiv \mathbbm{S}_{\tau}$ and compactify the 3-dimensional ball $\mathbb{R}^3_d$ to $\mathbbm{S}^3_z$.

\begin{figure}
	\centering
	\includegraphics[width=150mm]{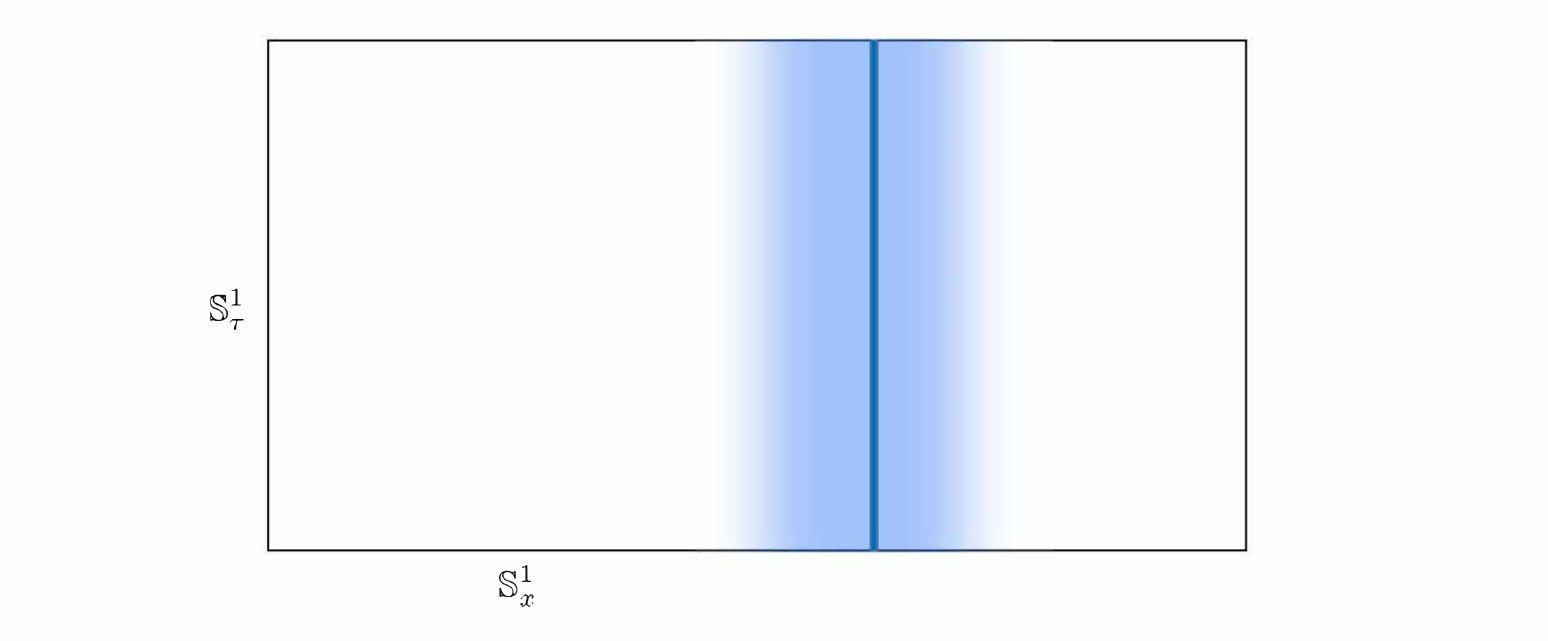}
	\caption{$\mathbbm{S}^1_x\times\mathbbm{S}^1_\tau$ part of the mapping torus $\hat{\mathbbm{Y}}_5$ where each point on the rectangle sits an $\mathbbm{S}^3_{z}$. The blue line stands for the location of a unit Yang-Mills instanton on $\mathbbm{S}^1_x\times\mathbbm{S}^3_{z}$ transported along $\mathbbm{S}^1_{\tau}$.
The twisting along $\tau$ direction together with the $\pi_3$ winding over  $\mathbbm{S}^3_{z}$ responsible for
the instanton induces a $\pi_4$ twisting as we glue the left end and the right end of this figure
to make $\mathbbm{S}^1_x$ and the mapping torus thereof. }
\end{figure}

We are then led to consider the mapping torus
\bea
\hat{\mathbbm{Y}}_5\; =\;\mathbbm{S}^1_x\times\mathbbm{S}^1_\tau\times\mathbbm{S}^3_{z}
\eea
that glues the vanilla $Y^{(x=0)}= \mathbbm{S}^1_\tau\times\mathbbm{S}^3_{z}$ to
the same manifold $Y^{(x=1)}= \mathbbm{S}^1_\tau\times\mathbbm{S}^3_{z}$ twisted
by the large gauge transformation $U(z,\tau)$ on the gauge bundle, a representative
of the nonzero element of $\pi_4(Sp(1))$. The key is that, given the unit $\pi_3$ winding of $U(z,\tau)$ for
each and every $\tau$, $\mathbbm{S}^1_x\times\mathbbm{S}^3_{z}$ part of the mapping torus
must have a $d=4$ unit instanton on it again for every value of $\tau$.

Furthermore, the twisting  $e^{\ii \pi\sigma_3}$ along $\mathbbm{S}^1_\tau$ acts
like $(-1)^F$ for half-integral isospin fermions, with $\tau$ considered
as the Euclidean time, so the mode analysis of fermions on $\hat{\mathbbm{Y}}_5$
comes with a periodic boundary condition along $\mathbbm{S}^1_\tau$.
Then, the fermion zero modes of the instanton in $\mathbbm{S}^1_x\times\mathbbm{S}^3_{z}$
elevates to fermion zero modes on the entire mapping torus
$\hat{\mathbbm{Y}}_5 =\mathbbm{S}^1_x\times\mathbbm{S}^1_\tau\times\mathbbm{S}^3_{z}$ by forgetting
the $\tau$-dependence. This connects the zero mode counting, mod 2, on the
$d=5$ mapping torus to the zero mode counting, mod 2, of a $d=4$ unit instanton.
In principle, the fermions with even isospin should be treated separately. Since they are known to contribute an even number of zero modes on a unit
instanton, one hopes that they would be irrelevant for the current discussion
of mod 2 counting here.

The same idea should work for all of $Sp(k)$ by embedding $Sp(1)\hookrightarrow Sp(k)$.
Since the number of
fermion zero modes is counted by the  Dynkin index $D_2^{\mathbf R}$.
With $Sp(1)=SU(2)$, for example, we have the Dynkin index of the isospin
$s$ representation, i.e., the rank $k=2s$ symmetric tensor,
\bea
D_2^{\boldsymbol\{2s\boldsymbol\}}(\ksp(1)) = \frac23 s(s+1)(2s+1)
\eea
so that each isospin $s\in 2\iZ+1/2$ fermion yields an odd number of zero modes for the
unit instanton.
For $Sp$ theories, the Witten anomaly will manifest, rendering the theory inconsistent, unless
the chiral fermion fields are collected to even out these zero-mode counts
\bea
\sum D_2^{\mathbf R} \;\in\; 2\iZ \ ,
\eea
giving us a simple litmus test against the Witten anomaly.

\subsection{Why Are $SU(N)$ Field Theories Safe?}

On the other hand, this still leaves a quandary, since the odd number of
fermion zero modes for a unit instanton is not an exclusive property of
$Sp$ theories. As explained in Appendix A, the above Pontryagin-Thom
construction fails to give a topologically robust $\pi_4$ configuration
for $SU(N)$, which is of course expected on account of $\pi_4(SU(N))=\emptyset$.

The direct connection between the mod 2 index of the mapping torus and
the instanton zero mode count mod 2 is thus no longer there.
Although we used an instanton over the mapping torus to motivate
the connection between $\pi_3$ and $\pi_4$ in the case of $\ksp$ theories,
the zero mode counting does not care whether it sits on
$\mathbbm{S}^1_x\times\mathbbm{S}^3_{z}$
or on  $\mathbbm{S}^1_\tau\times\mathbbm{S}^3_{z}$. The latter would
contribute to a purely four-dimensional instanton vertex, potentially plagued by
the fermion zero mode counting if the latter is odd.

A unit $Sp(1)=SU(2)$  instanton admits a single zero mode of the fermion
in the defining representation, for example, which extends to both $Sp(k)$
and $SU(N)$. This suggests that, despite the handy connection between the
two sides in the case of $Sp(k)$ gauge theories, the two problems associated
with $\pi_4$ and $\pi_3$ might be a priori two independent issues. With
$\pi_4(SU(N\ge 3))=0$ and the instantons with an odd number of fermion zero modes,
some explanation is necessary as to why $d=4$ $SU(N)$ theories are safe from the
latter's potential disease. For vanilla field theory in $d=4$, the resolution
comes from the cancelation of the perturbative gauge anomaly.

Unlike $Sp$ theories,
$SU$ theories may admit non-trivial anomaly polynomial \cite{Alvarez-Gaume:1983ihn},
\bea
P_6(\cF)\quad\sim\quad \sum_{\mathbf R} \pm \frac{1}{24\pi^2}
{\rm tr}_{\mathbf R} \cF\wedge\cF\wedge\cF
\eea
where $\pm$ again refers to the chirality of the fermion, which we
need to keep track of correctly for this perturbative anomaly.
In order to talk about the path integral, these perturbative anomalies
must cancel among themselves, and this prerequisite puts a
constraint on the chiral field content.
One can show for all possible $d=4$ $SU(N)$ theories that $P_6(\cF)=0$
enforces the net number of fermion zero modes in a unit $SU(N>2)$
instanton to be always even.

The perturbative $\ksu$ anomaly in $d=4$ is dictated by the sum,
\bea
\sum \pm \frac{1}{24\pi^2} \;{\rm tr}_{\bR} \cF\wedge\cF\wedge \cF
\eea
over the same set of fermions. Note from the above discussion of the anomaly
polynomials and the trace formulae, we always have
\bea
{\rm tr}_{\bR} \cF\wedge\cF\wedge \cF =C_3^{\bR} \times {\rm tr}_{\kg} \cF\wedge\cF\wedge \cF
\eea
with ${\rm tr}_{\kg}={\rm tr}_{\rm defining}$ for $\ksu$'s, so that
\bea
0=\sum_{\bR} \pm  C_3^{\bR}
\eea
is the perturbative anomaly-free condition.

For a minimal instanton, the fermionic zero modes are counted by
\bea
\sum\pm  D_2^{\mathbf R}(\kg)
\eea
where $D_2^\bR$ is the Dynkin index with $D_2^{\rm adj}=2h^\vee(\kg)$.
We saw earlier that only for $\ksu$'s and $\ksp$'s, odd values of
$ D^\bR_2$ are possible. Subtracting this last null expression, $\sum C_3^\bR$
from the sum of fermion indices above, we find a sufficient condition
for the net even number of fermion zero modes,
\bea
D_2^{\bR} - C_3^{\bR} \quad\in\quad 2\iZ
\eea
for all $\bR$.

For the fundamental representations i.e., for each of
rank-$k$ anti-symmetric tensors $\bR= \boldsymbol[k\boldsymbol]$ with
$k=1,2,\cdots, N-1$, we find
\bea
D_2^{\boldsymbol[ k\boldsymbol]} =\left(\begin{array}{c} N-2 \\ k-1\end{array}\right)\ , \qquad
C_3^{\boldsymbol[k\boldsymbol]}  =
\frac{N-2k}{N-k-1}\cdot \left(\begin{array}{c} N-3 \\ k-1\end{array}\right)
\eea
with which, after a simple manipulation, one finds
\bea
D_2^{ \boldsymbol[k\boldsymbol]} - C_3^{\boldsymbol[k\boldsymbol]}
&= & 2\times \left(\begin{array}{c} N-3 \\ k-2\end{array}\right)\quad\in\quad 2\iZ
\eea
always even, as promised. Using these iteratively and how $D_2$ and $C_3$
decompose identically under a tenor product, it is straightforward
to show that the same even-ness holds for all higher representations of $\ksu$'s.

We saw how the potential problem with an odd number of fermion zero modes in the
instanton sector is evaded by $SU(N)$ theories when the spectrum is free of the
perturbative one-loop anomaly. Although this removes the discrete inconsistency
for $d=4$ $SU(N)$ field theories, it also shows that the connection between
the Witten anomaly and its instanton litmus test should be taken more cautiously.

\subsection{Are $SU(N)$ Theories with Anomaly Inflow Safe?}

Now that we understand how the Witten anomaly and the instanton zero mode
counting interplay with each other in $d=4$ field theories, let us now turn
to a similar question in the context of string theory. What happens if $d=4$ 
$SU$ theory is one-loop anomalous but rendered consistent
because it is realized as part of a higher-dimensional set-up
such as superstring theory that supplies a canceling anomaly inflow?
The anomaly inflow by itself does not affect the anomalous $d=4$ fermion
content, so the above mechanism that evades the odd number of fermion
zero modes does not work anymore, leaving us a question of whether
and how instanton zero mode issue may be evaded in this higher
dimensional settings.

Superstring theories 
offers ample environment where we can embed chiral gauge theories as
a decoupling limit. 
One of the more prevalent such examples can be found in geometrical engineering 
in type II string theories\cite{Klemm:1996bj, Katz:1996fh}, or more precisely
in fractional D3 world-volume theories that probe the orbifold $\mathbb{C}^3/\Gamma$ with $\Gamma$ a discrete abelian subgroup of $SU(3)$\cite{Douglas:1997de}. More examples of such are about fractional D3 branes  probing toric Calabi-Yau 3-fold\cite{Feng:2000mi}. The resulting local Calabi-Yau's
come with natural quivers that can be inferred from the toric data, which can
be used either as the BPS quiver for $d=5$ Seiberg-Witten on a circle
when one starts from M-theory on a circle times the Calabi-Yau\cite{Morrison:1996xf, Intriligator:1997pq},
or for constructing (fractional) D3-probe field theories that
explore the geometry\cite{Hanany:2005ve,Franco:2005rj,Hanany:2006nm}.

One can see that a problem of the above kind with an odd number of instanton
zero modes can arise here when the net D3
charge of the probe is not integral. The fractional D3-branes are really
combinations of D5-branes and D7-branes wrapped on 2-cycles and 4-cycles,
respectively. The integral D3-brane corresponds to identical ranks
assignment to all nodes, in which case the incoming and the outgoing
arrow, or chiral and anti-chiral matter fields even out for any
of the quiver nodes. If we assign non-identical ranks to the nodes,
this translates to fractional D3-branes and potentially chiral matter spectrum.

\begin{figure}
	\centering
	\includegraphics[scale=0.8]{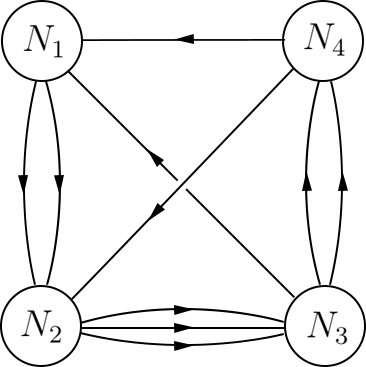}
	\caption{The BPS quiver for F1 theory}
\end{figure}

The simplest example of this is the so-called F1 geometry, a toric
Calabi-Yau that asymptotes to a conical geometry. The relevant quiver
comes with four nodes with one, two, and three arrows between nodes\cite{Closset:2019juk}.
With a judicious choice of the rank assignments to the four nodes,
therefore, we can construct $[\prod U(N_i)]/U(1)$ gauge theories
with various bifundamentals chiral fields which result in one-loop
gauge anomalies on $d=4$ intersections of such D5's and D7's.
The one-loop anomaly would be canceled by the universal I-brane anomaly inflow
\cite{Green:1996dd, Cheung:1997az, Minasian:1997mm, Kim:2012wc}
while $\pi_4$ anomaly is absent altogether.

Nevertheless, the question surrounding the odd number of fermion
zero modes that accompany the unit instanton is still
there. For example, this would be the case for $SU(N_1)$ instanton if
$N_3+N_4$ is odd, or the same for $SU(N_3)$ instanton if $3N_2-N_1$ is odd.
The resulting chiral matter content would generically generate perturbative
one-loop anomaly. Since the entire setup is embedded into a string theory,
however, the anomalies are canceled by some inflow mechanism. For the
examples at hand, the so-called I-brane anomaly inflow due to the
topological Chern-Simons-like couplings on D-branes does the job.

The setup is such that one cannot say that
the low energy effective theory is a $d=4$ gauge theory, since without the
anomaly inflow the $d=4$ path integral would be ill-defined. So this is not
quite $d=4$ field theory question. Nevertheless, the general
internal consistency of superstring theory suggests that the latter problem
should be also resolved in the end, with the likely answer being that additional
zero modes are generated from higher dimensional fermions in the set-up and
even out the $d=4$ problem. Exactly how this happens remains elusive to the authors.

Given this impasse, we would like to explore a pure field theoretical
venue where we can construct $d=4$ $SU(N)$ chiral theories purely in the field theory 
context but embedded in another pure field theoretical set-up in one higher
dimension. The one-loop anomaly is not canceled by  the spectrum but by an 
anomaly inflow, coming from the fact that $d=4$ theory is realized as a boundary 
theory of a highly gapped $d=5$ theory. This construction offers a simple 
laboratory where the instanton zero mode question of the above kind may be 
addressed in a self-contained manner.

\section{Witten-Yonekura Construction}

As we noted  at the end of the previous section, a new question arises when
we consider a chiral $SU(N)$ theory equipped with a perturbative gauge 
anomaly inflow. For $d=4$ chiral theories which are realized as part of some
higher-dimensional setup and which are rendered consistent at the perturbative 
level only thanks to an anomaly inflow, the potential problem with the
instanton zero mode counting disconnects from the Witten anomaly, so that
$SU(N)$ theory could in principle be in danger.

In this section, we will address this question in the context of recent construction by Witten and Yonekura\cite{Witten:2019bou, Yonekura:2016wuc, Dai:1994kq, Witten:2015aba}.
The latter is designed to address the anomaly inflow/descent of
discrete anomalies such as the Witten anomaly or the parity
anomaly, purely in the context of field theory, albeit starting from
one higher dimension. In particular, a chiral theory in even
dimension results as a boundary theory of edge modes from
a bulk theory with an arbitrarily large mass gap. The same setup
also happens to be a simple prototype for the perturbative anomaly
inflow for the perturbative anomaly, so offers an ideal playground for the $SU$
question we face here.

As we reviewed earlier, the Witten anomaly comes about because, for
Weyl fermions, the path integral is ill-defined by itself since
there is no well-defined eigenvalue problem. The remedy was to
extend the Weyl spinor artificially to a Dirac spinor with the well-defined
eigenvalue problem, and then collect half of the eigenvalues to form the Pfaffian.
This odd procedure of computing the partition function by inventing
the Dirac fermion as a middle step can be elevated to a more physical
one with the above construction with $Y_{2n}=\partial X_{2n+1}$, which
is precisely the Witten-Yonekura construction where the Dirac $\Psi$
as a physical field on $X_{2n+1}$.

This new setup, equally applicable when the dimension $d$ is odd as well,
does not cure the anomaly but rather recast it by elevating the partition
function on $Y_{d}$ to be one on $X_{d+1}$ upon an infinitely
heavy limit of the Dirac fermion $\Psi_{X_{d+1}}$ therein
\bea
\cZ_{Y_d}\equiv \lim_{M_{d+1}\rightarrow\infty} \mathbf Z_{X_{d+1}} = |\cZ_{Y_d}|\,e^{\ii \pi \eta^{\rm APS}(X_{d+1})}
\eea
where $\eta^{\rm APS}(X_{d+1})$ is the eta-invariant on $X_{d+1}$ computed with the
APS boundary condition on the boundary $Y_d$.
The point here is that with the appropriate sign choice of the mass $M_{d+1}$, the infinite
mass limit lifts all degrees of freedom in the bulk, except a single
massless chiral field $\psi_{Y_d}$ on $Y_d$ as a boundary degree of freedom.
See Appendix B for further details. When $d$ is odd, an immediate difference
is that both $\Psi$ and $\psi$ are considered Dirac fermions in the
respective dimensions.

\subsection{The Witten Anomaly as a Bulk Ambiguity}

Given $Y_d$, there must be more than one way to extend this to the bulk. Are different possible
choices of $X_{d+1}$ problematic? Let us consider a different extension
$X_{d+1}'$ such that $Y_d=\partial X_{d+1}'$, whereby the
potential ambiguity is
\bea
\frac{\mathbf Z_{X_{d+1}} }{\mathbf Z_{X_{d+1}'}} = e^{\ii \pi \eta^{\rm APS}({X_{d+1}})}e^{-\ii \pi \eta^{\rm APS}(X_{d+1}')} =    e^{\ii \pi \eta^{\rm APS}({X_{d+1}})}e^{\ii \pi \eta^{\rm APS}(-X_{d+1}')}
\eea
where $-X_{d+1}'$ means $X_{d+1}'$ with the orientation reversed. Also one should be mindful
that by the difference between ${X_{d+1}}$ and $X_{d+1}'$ we refer not only to the
geometry but to the gauge bundles over them as well. In fact, for the present context
of $d=4$, the primary interest would be in the difference of the gauge bundles.

The eigenmodes on ${X_{d+1}}$ and $-X_{d+1}'$ contributing to each eta invariant
are well localized in the bulk part of the respective manifold,
given how the APS boundary condition imposes an exponential decay
along the asymptotic cylinder of the boundary. This suggests that
the above ratio reduces to
\bea
e^{\ii \pi \eta^{\rm APS}({X_{d+1}})}e^{\ii \pi \eta^{\rm APS}(-X_{d+1}')} = e^{\ii \pi \eta({X_{d+1}}\cup(-X_{d+1}'))}
\eea
where the boundary condition is no longer needed  on the connected
sum ${X_{d+1}}\cup(-{X_{d+1}}')$, as the latter is a compact manifold.

The statement is then that there can be a potential ambiguity
in defining $\cZ_{Y_d}$ this way if for some compact manifold $\hat X_{d+1}$,
the eta invariant $\eta(\hat X_{d+1})$ is not an even integer, $\eta(\hat X_{d+1})\notin 2\iZ$.
Given such a $\hat X_{d+1}$, we can
always divide it by half with a shared boundary $Y_d$, on which $\psi$ lives,
such that
\bea
\hat X_{d+1}= X_{d+1}\cup(-X_{d+1}')
\eea
What would be the possible value of such $\eta(\hat X_{d+1})$?
With $d$ odd, $\eta(\hat X_{d+1})$ is in general an integer, equal to
the number of zero modes of $\ii\boldsymbol\gamma^A \sD_A$, since in even dimensions
the chirality operator flips the eigenvalue sign, so positive and negative
eigenvalues are matched 1 to 1.
What is a little surprising is that, although $\eta(\hat X_{2n+1})$
is not quantized for $d$ even, the collection of such $\eta$'s
from all the fermions is again an integer.

For a field theory on $Y_{2n}$ to make sense on its own, the cancelation
of the perturbative one-loop anomaly is necessary. In the absence of anomaly inflow,
this means \cite{Alvarez-Gaume:1983ihn} that
\bea
0=\left(\sum_{\rm chiral\;fields} \mathbb A\wedge {\rm ch}\right)\;\Biggr\vert_{d+2=2n+2}
\eea
Given several choices of the bulk extension $Y_{2n}= \partial X_{2n+1}= \partial X_{2n+1}'= \partial  X''_{2n+1}
= \partial X_{2n+1}'''$, imagine that $\hat X_{2n+1}= X_{2n+1}\cup(-X_{2n+1}')$ and  $\bar  X_{2n+1}=X_{2n+1}''
\cup(-X_{2n+1}''')$ are connected to each other continuously by ${\mathbbm X}_{2n+2}$, say,
$
\partial \mathbbm{X}_{2n+2}= -\hat X_{2n+1}+\bar X_{2n+1}
$.
The APS index theorem, together with the one-loop anomaly cancelation implies that
\bea
 \cI_{\rm APS}(\mathbbm{X}_{2n+2}) \;=\; \sum_{\rm chiral\;fields}\frac{\eta(\hat X_{2n+1})}{2}\;\;- \sum_{\rm chiral\;fields} \frac{\eta(\bar X_{2n+1})}{2}
\quad\in\quad\iZ
\eea
Equivalently,  we have
\bea
\prod_{\rm chiral\;fields} e^{\ii \pi \eta(\hat X_{2n+1})} \quad =\quad \prod_{\rm chiral\;fields} e^{\ii \pi \eta(\bar X_{2n+1})}
\eea
since  $e^{\ii 2\pi \cI_{\rm APS}(\mathbbm{X}_{d+2})}=1$.

For physical theories with the one-loop anomaly absent,
the potential ambiguity with how we extend $Y_{2n}$ to $X_{2n+1}$
is captured by collecting
\bea
\prod_{\rm chiral\;fields} e^{\ii \pi \eta(\hat X_{2n+1})}
\eea
for all possible compact manifolds $\hat X_{2n+1}$. The product is
invariant under arbitrary continuous deformations, so this collection
of phases is valued in the so-called cobordism group.

The Witten anomaly arising from $\pi_4(Sp(k))=\iZ_2$ suggests that
this phase ambiguity is at most $\iZ_2$-valued, or equivalently,
\bea
\sum_{\rm chiral\;fields}  \eta(\hat X_{2n+1})\quad\in\quad \iZ
\eea
If this sum is an odd integer, we would have $(-1)^N=-1$ in the
original language of the Witten anomaly, obstructing the Weyl fermion
partition function. In fact, this sum mod 2 computes the mod-2 index we
relied on when discussing the Witten anomaly via the mapping torus.

This reconstruction of the Witten anomaly via the bulk extension
may be considered the analog of the anomaly descent for the Witten 
gauge anomaly, a mathematical repackaging of the anomaly via a higher
dimensional ambient. One might also consider it an inflow, in that
it is a physical realization of the descent although not in the
sense of an anomaly cancelation mechanism.
The distinction would be whether one takes $X_{2n+1}$ and the
massive fermion theory over it as mathematical inventions or
physical entities. Next, we will turn to the latter viewpoint
in the context of $SU(N)$ theories.

\subsection{$SU(N)$ Theories with Anomaly Inflow}

In the above, we have assumed that the chiral theory on $Y_{2n}$
is free of the perturbative one-loop anomaly by itself, which allows
us to view the extension to $X_{2n+1}$ as a convenient mathematical
device for defining the partition function. However, we have learned
from many string theory realizations that a higher dimensional embedding
could be a physical reality. Part of the motivation behind the
Witten-Yonekura realization must have come from the topological
insulators where the topological bulk is the reality. We would
like to expand the discussion of anomaly a little here by considering
a model that gives a one-loop anomalous theory on $Y_{2n}$, only
to be canceled by an inflow from $X_{2n+1}$, and
see what happens to the discrete gauge anomalies thereof.

Consider an $SU(N)$ theory on $X_{2n+1}$ with $N>2$ and a single
boundary $Y_{2n}$ located at a large positive value of $x$.
Introduce $d=5$ massive fermion $\Psi$ in the complex
representations $\bR$ of $SU(N)$, which results in chiral $\psi_Y$'s such that
the field theory on $Y_{2n}$ is $SU(N)$-anomalous. However, if we view
the theory on $X_{2n+1}$ which comes with an arbitrarily large gap in bulk modes
as the definition of the theory in question, we have seen how
\bea
\mathbf Z_{X_{2n+1}} \;=\; |\cZ_{Y_{2n}}| \,e^{\ii \pi \eta_{\rm total}^{\rm APS}(X_{5})}\ ,\qquad \eta^{\rm APS}_{\rm total}(X_{5})=\sum_\bR\eta^{\rm APS}_\bR(X_{5})
\eea
can be taken as the definition of $\cZ_{Y_{2n}}$. The eta invariant,
the only phase factor here, is by definition a gauge-invariant
quantity, so this partition function is perfectly invariant under
the ``small" gauge transformation.

How should we understand this, given that the perturbative one-loop anomaly
results from the chiral spectrum? With the general relation between the eta
invariant in odd dimensions and the Chern-Simons term, the partition function
may also be written as
\bea\label{new_split_YW}
\mathbf Z_{X_{2n+1}} \;= \; \cZ'_{Y_{2n}} \,e^{ \ii \int_{X_{2n+1}} \sum\Omega_{2n+1}^{\rm total}(\cA)}
\eea
where $\cZ'_{Y_{2n}}$ would have resulted from the fully $d=4$ path integral
of $\psi_Y$ and contain the expected one-loop anomaly. The equality is exact as far as the
phase part goes. We have effectively split the eta invariant on $X_{2n+1}$
to the local and bulk part and the anomalous and non-local boundary
part.

The one-loop anomaly in $\cZ'_{Y_{2n}}$, computed by an
anomaly descent from some $P_{2n+2}(\cA)$, must be entirely canceled by an
inflow due to $\Omega_{2n+1}^{\rm total}(\cA)$; the two should be connected as
$P_{2n+2}(\cA)\sim -d\Omega_{2n+1}^{\rm total}(\cA)$. There can be in principle a
third piece-wise constant piece in the bulk, which usually accounts
for the difference between the gauge-invariant eta and the Chern-Simons action that
shifts under a large gauge transformation
in the case of a closed manifold. However, the latter becomes irrelevant
once exponentiated with the integer coefficient.

With this setup, let us revisit potential inconsistencies
associated with large gauge transformations and explore how the answers are
modified when the spectrum on $Y_{2n=4}$ is chiral and one-loop anomalous.
The Witten anomaly associated with $\pi_4(G)$ is absent, yet
we wish to explore the related observations about eigenvalue-crossings
and instanton zero-mode countings, and how the potential problem from
these are evaded here. It would be instructive to understand,
for example, how the Pontryagin-Thom construction works out for $SU(N)$,
or rather fails to generate a mapping torus, and how the $\pi_3(SU(N))=\iZ$
instanton physics is modified when we lift the inconsistent $Y_4$ theory
to a consistent $X_5$ theory.

For the eigenvalue-crossing, recall that the $d=4$ eigenvalues occur only
in pairs $(\lambda,-\lambda)$ for fermions in a complex gauge representation,
which potentially results in an odd number of eigenvalue-crossings when we connect
a pair of $Y_4$ via a five-dimensional cylinder. This could flip
$\cZ'_{Y_4}$ to $-\cZ'_{Y_4}$, in an apparent similarity to Witten's
$Sp(k)$ anomaly. In fact by embedding the of $Sp(1)=SU(2)$ mapping
torus, built up from Pontryagin-Thom construction, into $SU(N>2)$, one
can see easily that there would be a single eigenvalue crossing from
a single fundamental $\Psi$.

This may happen although the initial and the final
$Y_4$ would be the same, given $\pi_4(SU(N))=\emptyset$. However,
the point is that such a sign flip is no big deal for a one-loop anomalous
$SU(N)$ theory with its nontrivial and gauge-dependent phase despite how
the partition function was supposed to be a product of positive $\lambda$'s.
The best fix for such an anomalous phase here is to use $\mathbf Z_{X_5}$
as the definition of the gauge-invariant partition function, but the
latter is still equipped with a phase, albeit gauge-invariant now.

With a choice of $X_5$, the perturbative anomaly may be canceled out
at the level of $\bZ_{X_5}$ but the price we pay is that the partition
function becomes complex and the phase depends on which $X_5$ we
use for the extension. Given such phase ambiguities captured by
\bea
\frac {\mathbf Z_{X_{5}}}{\mathbf Z_{X'_{5}}}=  e^{\ii \pi \eta_{\rm total}(\hat X_5)}
\ , \qquad \hat X_5\equiv X_5\cup(-X'_5)\ ,
\eea
an extra sign flip due to an eigenvalue crossing somewhere along $X_5$ or $X_5'$
is hardly an issue.

When $P_6(\cF)\neq 0$, $\eta_{\rm total}(\hat X_5)$ need not be discrete.
Whether or not such a phase difference $\pi \eta_{\rm total}(\hat X_5)$ is a problem of consistency
depends on whether $X_5$ (or $X_5'$) here is a physical entity or a mathematical artifact.
If the former, the bulk extension with  the gauge field therein
would be considered part of the theory as well, so the phrase, including
the sign flip due to eigenvalue crossings, is a feature associated with
the choice $X_5$  among many such, not an ambiguity.

The perturbative anomaly inflow occurs in a variety of different
manners, rather routinely in string theory constructions for example,
so in the end we should expect many $SU(N)$-type chiral theories,
with perturbative anomaly canceled out by an inflow from how the
theory is embedded to a higher dimensional model. For such
constructions, the analog of $X_5$ is clearly part of the model, so
the phase analogous to $e^{\ii \pi \eta^{\rm APS}_{\rm total}(X_{5})}$
is an odd fact of life that we live with, rather than an inconsistency,
similar to the parity anomalies in odd dimensions.

\subsection{Back to the Instanton Zero Mode Issue}

All of these hair-splitting still leave the other problem due to zero
fermion modes on the unit $SU(N)$ instanton on $Y_4$. Although the
discussion revolving around the Thom-Pontryagin construction involves
an instanton string on the mapping torus, stretched along physical 
time direction, the potential inconsistency due to the instanton zero modes,
delineated by Witten and Goldberg, is very much a problem in the original
$d=4$ spacetime. 

\begin{figure}
	\centering
	\includegraphics[width=150mm]{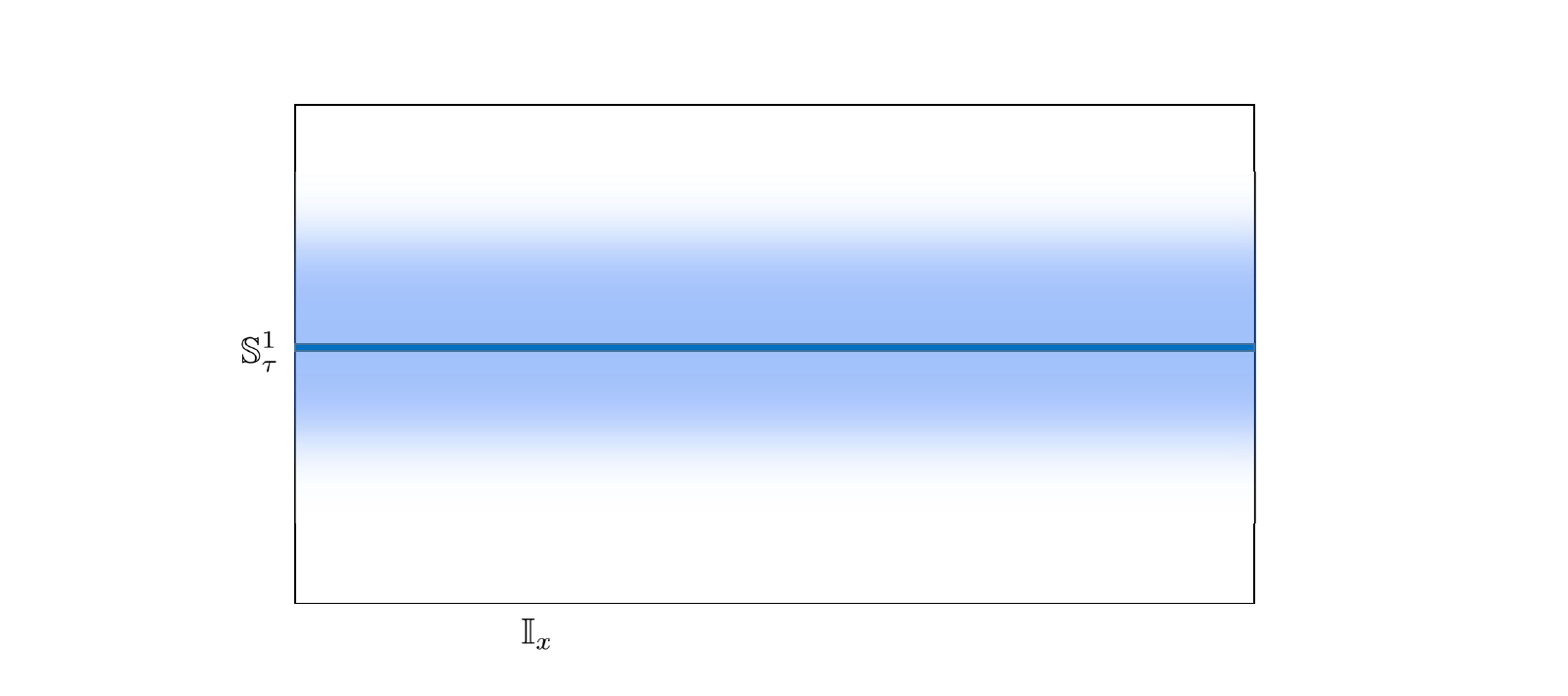}
	\caption{The configuration of a $d=4$ instanton embedded into the
Witten-Yonekura setup, now as an instanton string. We now consider a cylinder, 
$\mathbbm{I}_x\times\mathbbm{S}^1_\tau$, instead of mapping torus since the setup
needs to produce dynamical boundary fermions. This configuration would contribute to the path
integral as a nonperturbative saddle, playing the same role of $d=4$ instanton.
A boundary fermion of opposite chirality will emerge on the other end of the
instanton string, which also induces zero modes, doubling the latter in the end.
 }
\end{figure}

A new problem arises because we took care of the perturbative $SU(N)$ anomaly 
bluntly by an anomaly inflow rather than by carefully crafting the
fermion spectrum; the fermion zero mode counting around a $d=4$ instanton is
no longer constrained to be even. If we encountered a chiral theory with an odd
number of such fermionic zero modes, the non-perturbative instanton
vertex on $Y_4$ will be equipped with an odd number of fermions attached.
Does this not by itself lead to another inconsistency now related to 
$\pi_3(SU(N))=\iZ$? For $SU(2)$ we have seen how the presence of
such a problem coincides with that of the Witten anomaly from $\pi_4(SU(2))=\iZ_2$.

For the above Witten-Yonekura setup, thankfully, this last
concern also resolves itself naturally.
An instanton on $Y_4$ would elevate to a string of instanton
solutions that extend into the bulk $X_5$. There are three logical possibilities.

\begin{itemize}

\item $X_5$ closes off somewhere.

The instanton string with its non-trivial topology has nowhere to
end except back at $Y_4$ boundary, and would contribute as an instanton
and anti-instanton pair; individual instanton vertex with the problematic
odd number of fermion zero modes does not happen.
Although one could
imagine such a pair with a large mutual separation in $Y_4$, the action
would be at least proportional to the distance, being a string in $X_5$,
and exponentially suppresses the saddle; only the contributing pair would
be a tightly bound one where the worrisome odd number of flavor fermion
zero modes would be either doubled or more likely be lifted altogether.

\item
$X_5$ extends into an infinite semi-cylinder.

The instanton string with
a single end on $Y_4$ is possible but it would then stretch to the asymptotic
end and be of an infinite length and thus of infinite action, again suppressed
exponentially for the action.  This is further worsened by how the coupling
renormalization follows that of $d=5$ gauge theory, leading to arbitrary
weak coupling at the infrared end. No instanton vertex is possible in this case, either.

\item

$X_5$ is a finite semi-cylinder, more like $\mathbbm{Y}_5$, with two boundaries.

The line of Yang-Mills instanton stretched between the two ends is of finite length,
and the coupling renormalization turns over to that of $d=4$ as well. On the other
hand, we would find two boundary chiral theories at  two ends, say, at $x=0$ and $x=1$,
and the stretched $SU(N)$ instanton segment
would be equipped with fermion zero modes at each end. Because of the opposite $\boldsymbol\gamma^x $ chirality,
a zero mode of $\psi_{Y_4^{(0)}}$ would be matched by that of $\chi_{Y_4^{(1)}}^\dagger$
and vice versa; fermion zero modes are doubled, removing the instanton zero mode
quandary yet again.

\end{itemize}

Altogether, these take care of the potential consistency issue from the instanton
zero mode counting. We conclude that no new consistency issue
arises, despite that the fermion spectrum on $Y_4$ is one-loop
anomalous, as long as we view the theory on $X_5$ itself as a
physical entity.

\section{Soliton Zero Modes vs. Instanton Zero Modes}

Before closing, it is worthwhile to clear up potential confusion about
fermion zero modes, in how they enter the physics differently 
between the Lorentzian and the Euclidean settings. First, we need to 
emphasize more clearly that the fermionic zero modes we have been discussing
are complex ones. $d=4$ Weyl fermions are complex and the defining
representation of $SU(N)$ is also complex, so the zero modes would be also
complex. But this should raise an eyebrow. While we
made a big deal out of an odd number of zero modes, aren't these really 
even if we count them as real rather than as complex, since they are
accompanied by its complex conjugate? 

In the context of Lorentzian field theory, indeed, there is no issue
even if a nonperturbative object carries an odd number of complex fermionic 
zero modes. One typical example can be found with $SU(N)$ magnetic monopoles
in a theory equipped with fermions in the defining representation. For
the so-called fundamental monopoles \cite{Weinberg:1979zt}, the hypermultiplet in the defining
representation would either offer a single zero mode or none. In the 
former case, the low energy dynamics of the monopoles are well studied,\footnote{See
Ref.~\cite{Weinberg:2006rq} for a comprehensive review.} where
we find the quantum mechanics of such zero modes,
\bea
\int dt \;\ii \rho^\dagger \dot\rho+\cdots
\eea
where $\rho$ is the complex Grassman coefficient that multiplies the
$c$-number zero modes of the matter fermion in question. 

Quantized 
$\rho$ and its conjugate form a fermionic harmonic oscillator, so
this merely tells us about some internal degeneracy in the quantized
soliton, and further how the wavefunction of the soliton is a section
of a bundle over the soliton moduli space. 
In fact, depending on the gauge group and the matter representation
and also on the space-time dimension, one could imagine 
solitonic objects for which we may end up with a real version,
\bea
\int dt \;\frac{\ii}{2} \chi \dot \chi+\cdots
\eea
This generates a Clifford algebra of quantized $\chi$'s, and merely points
out that we need to think about a different bundle over the moduli space
as opposed to the case of complex $\psi$'s. There is no problem here when 
either the count of $\rho$'s or even that of $\chi$'s is odd.

In the context of the Euclidean signature, where we can now discuss
the nonperturbative saddles that contribute to the path integral, recall
how the fermion path integral is handled rather differently. Because
one cannot analytically continue $\bar\Psi=\Psi^\dagger\gamma^0$ to
the Euclidean side, we must invent an independent spinor $\Xi$ such 
that $\Xi^\dagger$ sits in place of $\bar\Psi$ in the field theory
action,
\bea
\int d^dx_L\; |e|\;\ii \,\bar\Psi \Gamma^a_L\mathscr D_a^L\Psi \qquad\rightarrow\qquad
\int d^dx_E\; |e|\;\ii\, \Xi^\dagger \gamma^a_E\mathscr D_a^E\Psi 
\eea
accompanied by the change of the path integral measure
\bea
[D\Psi^\dagger\,D\Psi]\qquad\rightarrow\qquad [D\Xi^\dagger\, D\Psi]
\eea
as well. The latter is analogous to how we can sometimes handle $d\bar z\, dz$
in the complex plane, pretending it to be $dw\, dz$ as the two are entirely
independent variables. Although sometimes this procedure is called the
fermion doubling, no actual doubling of the integration variable occurs.

In fact, this observation is a key to Fujikawa's derivation of
axial anomaly \cite{Fujikawa:1979ay}  and more generally is behind the Alvarez-Gaume and Witten's
treatment of all perturbative anomalies \cite{Alvarez-Gaume:1983ihn}. These anomalous phases arise because
$\Psi$ and $\Xi$ obey different mode-expansions in the zero-eigenvalue
sector when the fermions are coupled chirally,
\bea
\Psi=\sum_{\rm I} \rho^E_I\Psi_I^{(0)} +\cdots \ , \qquad \Xi=\sum_{p} \xi^E_p\Xi_p^{(0)} +\cdots
\eea
where $I$ and $p$ respectively label the kernels of the two spinors, distinct
from each other if these spinors are chiral. 

With generic gauge field that belongs to a topologically nontrivial gauge bundle
and with appropriately chiral $\Psi$, we often end up with
\bea
\Psi=\sum_{\rm I} \rho^E_I\Psi_I^{(0)} +\cdots \ , \qquad \Xi=\cdots
\eea
or vice versa, with the ellipsis denoting nonzero eigenmodes which pairwise 
match between the two spinors. The key insight by Fujikawa is that this
disparity in the integration measure accounts for the axial anomaly as viewed
from the Euclidean side.\footnote{There is more to Fujikawa's argument than this
since the axial anomaly is present regardless of topologically nontrivial gauge 
bundles. Instead, one also needs to take into account a related disparity between
the eigenvalue densities in the continuous non-kernel part of the spectrum as well so that it is
actually not the index but the so-called bulk index that determines the
anomaly. Our point here remains valid in that the eigenfunction spectrum 
of $\Psi$ and $\Xi$ should be treated as independent. }

With Yang-Mills instantons, then, we arrive at the usual story in how the
saddle contribution vanishes unless an appropriate collection of fermions
is inserted at the path integral. Or equivalently, the presence of the zero
modes implies that the saddle contribution, say in the dilute gas approximation,
can be emulated by inserting a vertex in the effective action of type
\bea
\int dX e^{-S^E(X)}\prod \Psi(x)
\eea
plus its conjugate term, where the number of $\Psi$ equals the number of 
zero modes $\Psi^{(0)}_I$. $S^E(X)$  is the Euclidean action of the instanton 
while $X$ are collective coordinates of  the instanton in question, among 
which are the spacetime position $x$ of the saddle configuration. 

One problem is that the odd number of  fermionic zero modes thus manifests 
here as an effective  vertex of the Grassman-odd type.
It is not entirely clear if the path integral prohibits such a vertex. For instance,
at the level of quantum fermionic harmonic oscillator, a term like that in the
Hamiltonian would be merely an off-diagonal piece that connects the bosonic 
and the fermionic states related by the raising/lowering operator. 
With field theories, however, there is an additional need for the effective 
vertex to be invariant under (Euclideanized) Lorentz transformation, which 
would be very difficult to meet for an odd number of $\Psi$'s. This latter observation 
complements the earlier inconsistency argument by Goldstone. 

Before closing we would like to mention that there is a similar issue in $d=3$
$SU(N)$ theory in the Coulomb phase, with fermions in the defining representation.
Here the fundamental monopoles act like an instanton due to one-less spacetime
dimensions, and, as noted above, depending on the mass of the matter fermion, 
one can easily find situations with an odd number of matter fermion zero modes 
attached to such Euclidean monopoles. This would cause trouble along the same lines as above.

Traditionally, this type of situation has been sidestepped by requiring 
the integer-quantized Chern-Simons level shift from matter one-loop for the
presumed consistency under large gauge transformations. However, the fermion
one-loop actually generates the eta-invariant \cite{Alvarez-Gaume:1984zst,Witten:2015aba}, which is entirely a gauge-invariant
object; the Chern-Simons piece is only the continuous part of this eta-invariant,
whose failure of be invariant under the large gauge transformations is corrected
by the piece-wise constant remainder in the eta-invariant. This
old prejudice against the odd number of $d=3$ matter fermion in the defining
representation of $SU(N)$, on account of proper quantization of Chern-Simons level, 
is not justified and is unable to preclude the quandary here.

Along with the instanton zero-mode problem in fractional D3 probe theories earlier,
we leave further investigation of these issues for a future study.

\section*{Acknowledgements}
We thank Amihay Hanany, Yang-Hui He, and Heeyeon Kim for useful discussions. This work is supported by KIAS individual grants, PG005705 (PY) and PG080802 (QJ).

\appendix

\section{Pontryagin-Thom  and Its Failure for $SU(N)$}
Consider $X$ as any compact manifold and $Y, Y'$ are two compact submanifolds. Then we say $Y$ and $Y'$ are cobordant within $X$ if there exists a compact manifold $Z \subseteq X \times [0,1] $ connecting $Y$ with $Y'$ such that,
\begin{equation}
    \partial Z = Y \times \{0\} \ \cup\ Y'\times\{1\}.
\end{equation}
There are many refinements of the basic notion of cobordism and the relevant here is the \textbf{framed} cobordism. A \textbf{framing} of a submanifold $Y$ is a smooth basis of section $\mathfrak{o}_Y$ of the normal bundle $N_{Y/X}$. Then two framed submanifolds $Y$ and $Y'$ are \textbf{framed cobordant} if there exists a cobordism $Z$ together with a smooth framing $\mathfrak{o}_Z$ of the normal bundle $N_{Z/X\times[0,1]}$ which restricts to $\mathfrak{o}_N$ and $\mathfrak{o}_{N'}$ at $Y\times\{0\}$ and $Y\times\{1\}$ respectively.

There is a simple homotopy description of the framed cobordism.

\begin{theorem}[Pontryagin-Thom]
    The equivalence classes of \textbf{framed} submanifolds of codimensional $r$ are one-to-one correspondence with $[M,S^r]$, where $[M,S^r]$ is the homotopy class of the map $M\rightarrow S^r$
\end{theorem}

\begin{figure}
    \centering
    \includegraphics[scale=0.5]{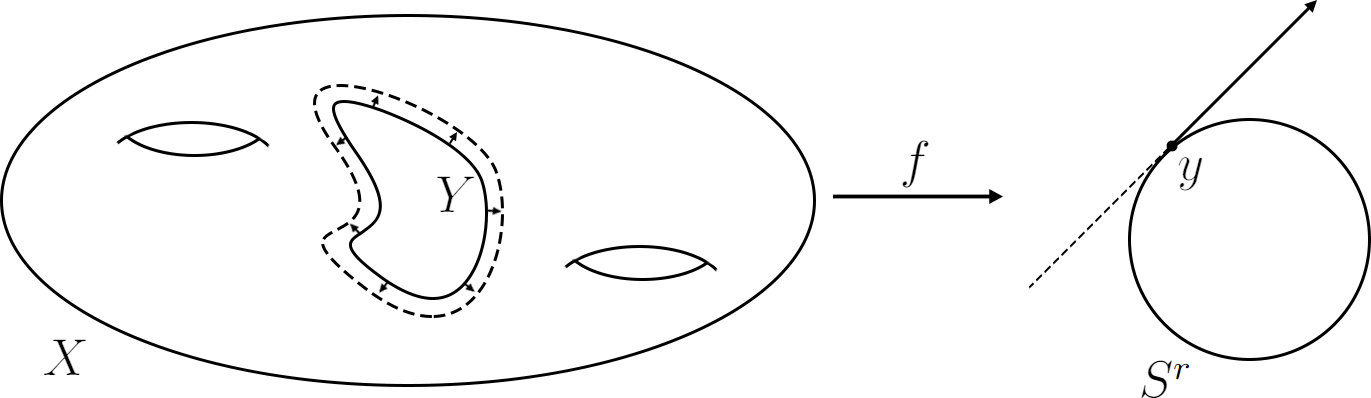}
    \caption{Homotopy map $f: X \rightarrow S^r$ and the submanifold $Y$.}
    \label{fig:PT-homotopy-map}
\end{figure}

One can find a review of this in the Appendix of \cite{freed1991instantons}. We will not try to prove the theorem in the following, instead, we will construct the correspondence explicitly. Consider a homotopy map $f: X \rightarrow S^r$ and a point $y\in S^r$, see figure \ref{fig:PT-homotopy-map} for an illustration. Then $Y=f^{-1}(y)$ is a smooth codimensional-$r$ submanifold in $X$. Then for each point $x\in N$, $df_x$ maps the normal space $TX_x/TY_x$ isomorphically onto the tangent space $T_yS^r$ at $y\in S^r$.

So fixing a frame $\mathfrak{o}_y$ at $y\subset S^r$, we obtain a framing of $Y$ by pullback:$\mathfrak{o}_Y(x) = (d f_x)^{-1}(\mathfrak{o}_y)$.

\begin{figure}
    \centering
    \includegraphics[scale=0.5]{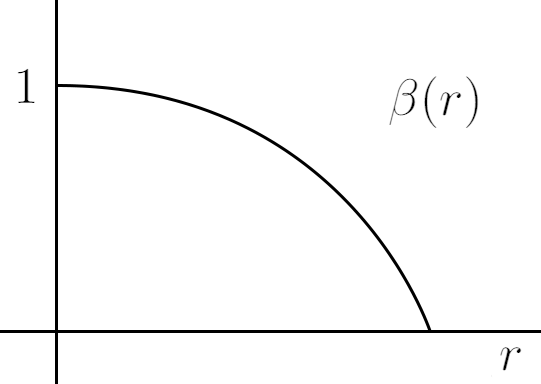}
    \caption{The cutoff function $\beta(r)$.}
    \label{fig:PT-cutoff-function}
\end{figure}

Conversely, if we have a framed codimensional-$r$ submanifold $Y$ in $X$, the framed normal bundle can be realized by a tubular neighborhood $Y\times \mathbb{R}_d^r\subseteq X$ where $\mathbb{R}^r_d$ is a $r$-dimensional ball with a small radius $d$ and the standard frame of $\mathbb{R}_d^r$ induce the section $\mathfrak{o}_Y$. We can build the homotopy map $f$ from $X$ to $S^r$ as follows. Introduce a cutoff function $\beta(r)$ with $r\geq 0$ as shown in figure \ref{fig:PT-cutoff-function} which is one near $r=0$ and decreases to zero rapidly as $r$ grows.
Define the projection map $\pi: Y \times \mathbb{R}^r\rightarrow \mathbb{R}^r$ and the homotopy map $f: Y \rightarrow S^r$ is constructed as,
\begin{equation}\label{PT-homotopy-map}
   f(x) = \left\{ \begin{array}{ll}
         \frac{\pi(x)}{\beta(|x|)}\quad \textrm{if}\ x \in Y \times \mathbb{R}_d^r \\
          \infty\quad \textrm{if}\ x \notin Y \times \mathbb{R}_d^r
    \end{array}\right.
\end{equation}
where $S^r$ is identified with $\mathbb{R}^r \cup \{\infty\}$ by stereographic projection. The homotopy class $[f]\in [X,S^r]$ is determined uniquely by $Y$ and its framing $\mathfrak{o}_Y$.

In section 3, we have illustrated the homotopy map corresponding to $\pi_4(S^3) = \mathbb{Z}_2$ as an example of Pontryagin-Thom's construction. Since the group manifold of $Sp(1)=SU(2)$ is $S^3$, the Pontryagin-Thom gives a construction of the large gauge transformation $[U(x)] \in \pi_4(Sp(1))$ on $S^4$. Naively, one may embed $SU(2)$ into $SU(N)$ ($N \geq 3$) and construct a similar gauge transformation on $S^4$ with gauge group $SU(N)$. However, due to the fact $\pi_4 (SU(N))=0$ ($N \geq 3$) the map $f(x)$ must be topologically trivial. In the remainder of this Appendix, we will clarify this point.

Without loss of generality, let's focus on the $SU(3)$ case. The group manifold of $SU(3)$ is 8-dimensional and locally $S^5 \times S^3$. The two factors can be understood as follows. Consider a $3\times 3$ dimensional unitary matrix $M \in SU(3)$ and focus on the first column,
\begin{equation}
    M = \left[\begin{array}{ccc}
        z_1&*&*\\
        z_2&*&*\\
        z_3&*&*
    \end{array}\right]
\end{equation}
which is a complex vector $(z_1,z_2,z_3)$ with unit norm $|z_1|^2 + |z_2|^2 + |z_3|^2 = 1$. It defines a unit 5-sphere $S^5$ which is the first factor. Then we can consider a unitary transformation and set $(z_1,z_2,z_3)$ to $(1,0,0)$ and $M$ becomes,
\begin{equation}
    M = \left[\begin{array}{ccc}
        1&0&0\\
        0&*&*\\
        0&*&*
    \end{array}\right]
\end{equation}
and the $2\times 2$ block inside $M$ is an $SU(2)$ matrix which corresponds to the second factor $S^3$. Therefore $SU(3)$ is locally $S^3 \times S^5$.

Globally $SU(3)$ is a fiber bundle with base $S^5$ and fiber $S^3$ constructed as follows. Split $S^5$ into two hemispheres $S^5_+$ and $S^5_-$ and they are glued along the equator $S^4$. The transition function between the two hemispheres is $t_{+-}$, which defines a map from the equator $S^4$ to $SU(2)$ matrix, where $SU(2)$ acts on the fiber $S^3$ by treating $S^3$ as another $SU(2)$ matrix and multiplying them together. Therefore $t_{+-}$ is classified by the homotopy group $\pi_4(S^3) = \mathbb{Z}_2$. If $t_{+-}$ corresponds to the trivial element in the homotopy group then the bundle is simply $S^5 \times S^3$; on the other hand, if $t_{+-}$ corresponds to the non-trivial element the bundle is $SU(3)$ manifold.

Now let's consider the map $f(x): S^4 \rightarrow S^3$ defined as \eqref{PT-homotopy-map} with $X=S^4$ and $Y=\gamma$ is a circle, which gives a non-trivial element in the homotopy group $\pi_4(S^3)$. We may identify $S^3 = SU(2)$ with the fiber of the $SU(3)$ manifold and lift $f(x)$ to a map from $S^4$ to $SU(3)$ which maps the whole $S^4$ to the $SU(2)$ fiber at the north pole of $S^5_+$. The question is whether $f(x)$ can be deformed to a trivial map or not.

\begin{figure}
    \centering
    \includegraphics[scale=0.5]{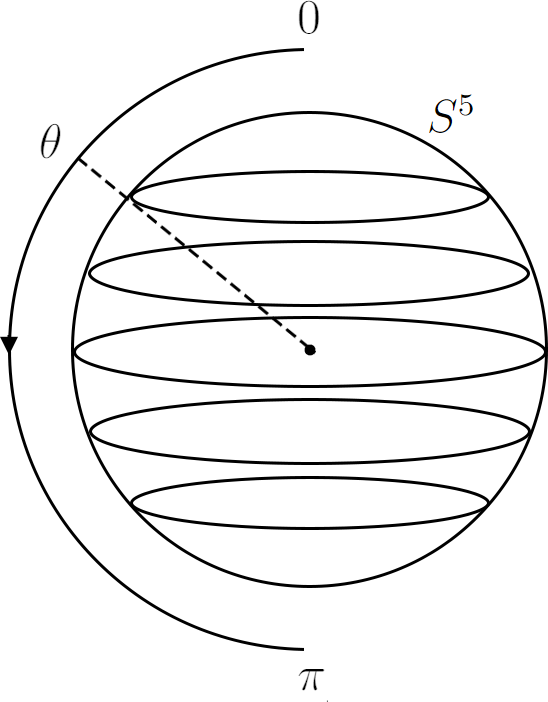}
    \caption{The family of $\widetilde{S}^4_{\theta}$}
    \label{fig:PT-S4-deformation}
\end{figure}

To answer this question, we resolve the north pole to a 4-sphere $\widetilde{S}^4_{\theta} \subseteq S^5$ centering the north pole and polar angle is controlled by $\theta \in [0,\pi]$ as illustrated in figure \ref{fig:PT-S4-deformation}.
When $\theta=0$ we recover the north pole and when $\theta=\pi/2$ the 4-sphere is the equator of $S^5$. For $\theta\geq \pi/2$ the 4-sphere moves to the south hemisphere and finally when $\theta=\pi$ the 4-sphere shrinks to the south pole.

Consider the local bundle $\widetilde{S}^4_{\theta} \times S^3$ inside $SU(3)$ and we can identify $\widetilde{S}^4_{\theta}$ with $X=S^4$ with an isomorphism $i(x)\in\widetilde{S}^4_{\theta}$ such that $f(x)$ induce a section of $\mathfrak{s}_{\theta}$ on $\widetilde{S}^4_{\theta} \times S^3$ satisfying $f(x) = \mathfrak{s}_{\theta} (i(x))$. Introduce the family of map $f_{\theta}(x) = \mathfrak{s}_{\theta} (i(x)) : S^4 \rightarrow SU(3) $ as a smooth deformation of the map $f(x)$. When $\theta=0$ all the fibers on $\widetilde{S}^4_{\theta}$ collide and one has $f_0(x) = f(x)$. On the other hand, $f_{\pi/2}(x) = \mathfrak{s}_{\pi/2} (i(x)) $ is a section of $\widetilde{S}^4_{\pi/2} \times S^3$ where $\widetilde{S}^4_{\pi/2}$ is the equator of $S^5$. We can further extend it into the south hemisphere $S^5_{-}$ by applying the transition function $t_{+-}$,
\begin{equation}
    \mathfrak{s}'_{\pi/2} (i(x)) = t_{+-} \circ \mathfrak{s}_{\pi/2} (i(x)).
\end{equation}
However, both $t_{+-}$ and $\mathfrak{s}_{\pi/2}$ belong to the non-trivial element of the homotopy group $\pi_4(S^3)=\mathbb{Z}_2$ which means  $\mathfrak{s}'_{\pi/2} (i(x))$ can be trivialized in the south hemisphere. Therefore when $\theta=\pi$, $f_{\pi}(x)$ becomes a trivial map that maps the whole $S^4$ to the south pole and we have shown that $f(x)$ can be deformed to a trivial map through $f_{\theta}(x)$.

\section{Witten-Yonekura}

We begin with a massive Dirac fermion $\Psi$ with mass $M$ on a $2n+1$-dimensional manifold $X_{2n+1}$. The Euclidean action is
\begin{equation}
    I = - \int_{X_{2n+1}} d^{2n+1} x \sqrt{g} \bar{\Psi} ( \ii \boldsymbol\gamma^{A} {\sD}_{A} + \ii M)\Psi,
\end{equation}
with $A=1,\cdots,2n+1$ and $\Psi$ is coupled to the Riemannian metric $g$ and background gauge field $\mathcal{A}$ on $X_{2n+1}$. The Dirac operator $\ii \boldsymbol\gamma^A {\sD}_A$ is Hermitian if $X_{2n+1}$ is a closed manifold without boundary. Suppose the manifold $X_{2n+1}$ has a single boundary $Y_{2n} = \partial X_{2n+1}$ and the metric near the boundary takes the form,
    \begin{equation}
        d s_{X_{2n+1}}^2 = d x^2 + d s_{Y_{2n}}^2,
    \end{equation}
where $x \in (-\epsilon,0]$ parameterizes the normal direction which vanishes along $Y_{2n}$ and is negative in the bulk. Impose the local boundary condition $\mathbf{L}$,
    \begin{equation}
        \mathbf{L} : \left( 1 + \boldsymbol\gamma^{x}\right) \Psi|_{x=0} = 0,
    \end{equation}
where $\boldsymbol\gamma^{x}$ is the gamma matrix along the $x$ direction and is also the chirality operator on $Y_{2n}$. In other words, the boundary condition $\mathbf{L}$ requires $\Psi$ to be chiral when restricting along the boundary $Y_{2n}$. The Dirac operator $\ii \boldsymbol\gamma^A {\sD}_A$ is no longer hermitian with such a boundary condition: one can perform an integration by parts and let the Dirac operator act on $\bar{\Psi}$ in the action, and it will produce an additional boundary integral which is non-zero with the $\mathbf{L}$ boundary condition.

The Dirac equation near the boundary can be written as,
	\begin{equation}
		\ii \boldsymbol\gamma^x \left( \partial_x + \mathbb D_{(x)} +  \boldsymbol\gamma^x M\right)\Psi=0, \qquad \mathbb D_{(x)}\equiv\boldsymbol\gamma^x \gamma^a{\sD}_a\biggr\vert_x
	\end{equation}
with $a = 1,\cdots,2n$. Solving the Dirac equation near the boundary gives,
\begin{equation}
    \Psi = \psi_{Y_{2n}} \exp(M x),\quad \left( 1 + \boldsymbol\gamma^x\right) \psi_{Y_{2n}}=0,\quad \mathbb D_{(x)} \psi_{Y_{2n}} = 0,
\end{equation}
where $\psi_{Y_{2n}}$ is a chiral Dirac fermion living at the boundary $Y_{2n}$ and it solves the $2n$-dimensional massless Dirac equation. If $M>0$, this mode decays exponentially when $x<0$ and is localized along $Y_{2n}$. On the other hand, if $M<0$ it spreads over $X_{2n+1}$ and is not normalizable. We also introduce a simple Pauli-Villars regulator which is a very massive fermion with mass $M_r$ of opposite statistics. Since we do not want the regulator field to have a low energy mode propagating along the boundary, we require the mass $M_r$ to be negative.

If we treat $x$ as the time direction, the partition function of the massive fermion can be written in Hamtionian formalism as,
    \begin{equation}
        \mathbf{Z}_{X_{2n+1}} = \langle \mathbf{L} | X_{2n+1} \rangle,
    \end{equation}
where $|X_{2n+1}\rangle$ is a state vector defined by the path integral over $X_{2n+1}$ and $\langle \mathbf{L}|$ is the state vector defined by the boundary condition. Both of them belong to the Hilbert space $\mathcal{H}_{Y_{2n}}$ on $Y_{2n}$. We will take the mass $|M|$ to be very large such that there is a large mass gap in the Hilbert space $\mathcal{H}_{Y_{2n}}$. Therefore all the massive modes are suppressed and the vector $|X_{2n+1}\rangle$ is proportional to the ground state $|\Omega\rangle$. Assuming $\langle \Omega | \Omega \rangle =1$, we can split the bulk and boundary contribution of the partition function as,
    \begin{equation}
        \mathbf{Z}_{X_{2n+1}} = \langle \mathbf{L}|\Omega \rangle \langle \Omega | X_{2n+1}\rangle.
    \end{equation}
However, both $\langle \mathbf{L}|\Omega \rangle$ and $\langle \Omega | X_5\rangle$ are not well-defined due to the phase ambiguity of the ground state $|\Omega\rangle$. If we denote $\mathcal{W}$ as the parameter space including metric $g$ and gauge background $\mathcal{A}$ on $Y_{2n}$, an adiabatic change represented as going along a loop in $\mathcal{W}$ may induce a nontrivial Berry phase to the ground state $|\Omega\rangle$. This ambiguity is related to the fact the partition function of a chiral fermion on $Y_{2n}$ is not well-defined\cite{Yonekura:2016wuc}. Assuming the boundary Dirac operator $\mathbb D_{(x)}$ has no zero mode, the partition function can be evaluated by inserting a pair of APS (Atiyah-Patodi-Singer) boundary states in between,
    \begin{equation}
        \mathbf{Z}_{X_{2n+1}} = \frac{\langle \mathbf{L}|\Omega \rangle \langle \Omega | \textrm{APS}\rangle}{|\langle \textrm{APS} | \Omega \rangle|^2} \langle \textrm{APS} | X_{2n+1}\rangle.
    \end{equation}
The APS boundary condition is defined by requiring $\Psi$, restricted to the boundary $Y_{2n}$, to be expanded as a linear combination of eigenmodes of $\mathbb D_{(x)}$ with only positive eigenvalues and all the eigenmodes with negative eigenvalues are projected out. It is equivalent to attaching a semi-infinite long cylinder $Y_{2n} \times [0,\infty)$ to $X_{2n+1}$ such that only the modes with positive eigenvalues are normalizable when $x \rightarrow +\infty$, which is the set-up of APS index theorem.

The Dirac operator $\ii \boldsymbol\gamma^{A} {\sD}_{A}$ is hermitian with respect to the APS boundary condition because the leftover at the boundary after the integration by part is zero. Therefore the eigenvalues $\{ \lambda \}$ of $\ii \boldsymbol\gamma^{A} {\sD}_{A}$ are real and one may write the term $\langle \textrm{APS} | \Omega\rangle$ as,
\begin{equation}
    \langle \textrm{APS} | X_{2n+1} \rangle = \prod_{\lambda} \frac{\lambda + \ii |M|}{\lambda - \ii |M_r|} = \left(\prod_{\lambda} \frac{\lambda + \ii |M|}{\lambda}\right) \left(\prod_{\lambda} \frac{\lambda}{\lambda - \ii |M_r|} \right),
\end{equation}
where we have included the regulator field $\Psi_r$ with mass $M_r$. Both $M$ and $M_r$ should be sent to infinity and we will set $|M_r|=|M|$ in the following. Notice that,
\begin{equation}
    \lim_{M \rightarrow \infty}\prod_{\lambda} \frac{\lambda}{\lambda \pm \ii M} = \left(\prod_{\lambda}\left|\frac{\lambda}{M} \right| \right) e^{\ii \sum_{\lambda} \textrm{Arg}\left(\frac{\lambda}{\lambda \pm \ii M}\right)} = \left(\prod_{\lambda}\left|\frac{\lambda}{M} \right| \right) e^{\mp \ii \frac{\pi}{2} \sum_{\lambda} \textrm{sgn}(\lambda)},
\end{equation}
where the summation $\sum_{\lambda} \textrm{sgn} (\lambda)$ is regularized using the $\eta$-invariant,
    \begin{equation}
        \eta^{\textrm{APS}}(X_{2n+1}) \equiv \lim_{s\rightarrow 0} \sum_{\lambda} \textrm{sgn}(\lambda) e^{-s|\lambda|}.
    \end{equation}
Therefore $\langle \textrm{APS} | X_{2n+1} \rangle$ can be evaluated as,
    \begin{equation}
        \lim_{M \rightarrow \infty} \langle \textrm{APS} | X_{2n+1} \rangle = \exp\left(\ii \pi \eta^{\textrm{APS}}(X_{2n+1}) \right).
    \end{equation}
The treatment of the rest term is more complicated and one needs to evaluate the overlap $\langle \mathbf{L} | \Omega \rangle$ and $\langle \textrm{APS} | \Omega \rangle$. They are computed by switching to the Lorentz signature and doing a straightforward quantization of the fermion $\psi_{Y_{2n}}$ on the space $Y_{2n}$. We refer to \cite{Witten:2019bou} for a detailed computation and the result is,
    \begin{equation}\label{YW-Partition-function-X5}
        \frac{\langle \mathbf{L}|\Omega \rangle \langle \Omega | \textrm{APS}\rangle}{|\langle \textrm{APS} | \Omega \rangle|^2} = |\cZ_{Y_{2n}}|
    \end{equation}
where $\cZ_{Y_{2n}}$ is the (anomalous) partition function of the chiral Dirac fermion $\psi_{Y_{2n}}$ living at the boundary $Y_{2n}$.

Combine the two terms, the total partition function of the bulk massive fermion $\Psi$ with the boundary condition $\mathbf{L}$ is given by,
\bea
\lim_{M\rightarrow\infty} \mathbf Z_{X_{2n+1}} = |\cZ_{Y_{2n}}|\,e^{\ii \pi \eta^{\rm APS}(X_{2n+1})}
\eea
Both two factors are separately gauge invariant but they are not separately physically sensible. Whenever the boundary Dirac operator $\mathbb D_{(x)}$ develops a zero mode near a value of background field $g$ and $\mathcal{A}$, neither two factors vary smoothly. The Dai-Freed theorem\cite{Yonekura:2016wuc, Dai:1994kq} ensures the product is smoothly varying.

\end{document}